# Coulomb-like Model for International Trade Flow and Derivation of Distribution Function for Trade Flow Strength

Mikrajuddin Abdullah

*Department of Physics, Bandung Institute of Technology, Jalan Ganesa 10, Bandung, 40132, Indonesia*

**Abstract**

To describe international trade flows, we propose the coulomb force formulation, in which the magnitude of the charge represents gross domestic product (GDP) and the distance between countries is the bilateral distance, the product of spatial distance and "dielectric constant," rather than the spatial distance as used in the gravitation model, allowing it to be time dependent. The "dielectric constant" is influenced by factors such as warfare, transportation disruptions, trade agreements, social, geography, politics, culture, and others. The GDP and distance power parameters were estimated using data from high-GDP countries' export-import transactions. We also developed a trade strength distribution equation that fits World Bank data reasonably well over a decade.

## I. INTRODUCTION

In economics, the theory of gravity has been used to explain trade flows between countries [1,2,3,4,5,6]. In the initial formulation, the strength of the gravitational force between two objects represents the trade volume between the two countries, the mass of the objects represents the GDP, and the distance between the objects represents the distance between the capital cities of the two countries [7]. However, additional corrections are required, such as by including the GDP/capita, trade agreement factors, random factor, GDP and distace power parameters, and possibly other factors [2,8,9,10,11] due to presence of gaps between theory and empirics [12].

By taking into account other factors such as social, geography, politics, and culture, the gravity model is consistent with any international trade [13,14] by considering the distance as a measure of the bilateral distance between the two countries [15], which does not always refer to physical distance. Warfare, disruptions in transportation routes, trade agreements, and other factors can all have an impact on such a distance. For example, European economic integrations such as the European Community and the European Free Trade Agreement have been shown to significantly increase inter-member trade flows [16,17]. Karemera and Koo [18] empirically assessed the effects of the US-Canada free trade agreement on trade expansion as bilateral distances between countries were reduced.

There are still several issues with the current gravity model. First, in the gravitational equation, the distance between the two countries is defined as the spatial distance between capital cities [7] or distance between major seaports or airports [19], which is a constant parameter that cannot accommodate a fluctuating "bilateral distance" [13]. Second, the GDP of countries around the world ranges from a few million USD to more than 20 trillion USD [20], and there has been no report on the derivation of the GDP distribution equation, beginning with the most fundamental formulation. Third, the gravity model is more frequently discussed on the economic side, which places more emphasis on estimating the power parameters of the GDP and distance [2,7,9,21,22,23,24] and lacks exploration of physics concepts [19,25].

Another natural law that explains the force between objects is Coulomb's law which has the same form as the law of gravity, but there is an effect of the dielectical constant of material. To return to the gravity/coulomb equation's into original form, we can incorporate all of the factors influencing trading volume into the “dielectric constant” parameter. The bilateral distance is defined as the product of spatial distance and the “dielectric constant”. When there is tension between two countries, the “dielectric constant” between them approaches infinity, and there is no trade flow between them despite the fact that the distance between the two capital cities is short. In contrast, if two countries have a trade agreement and no barriers exist, the “dielectric constant” is a minimum.

The goal of this paper is to explain international trade flow between countries using the coulomb force formulation. Then, adopting the derivation of gravitational fluctuations discussed by Chandrasekhar and von Neumann [26], we will derive for the first time the distribution equation of trade flow strength between countries.

## II. METHOD

The trade flow equation between countries $m$ and $n$ is written in the form of the coulomb equation (see explanation in **Supplementary 1**)

$$F_{mn}(t) = K\frac{G_m^\delta G_n^\delta}{T_{mn}^\beta} \quad (1)$$

where $T_{mn} = \omega_{mn}^{1/\beta} R_{mn}$ denotes the bilateral distance between the two countries, $R_{mn}$ denotes the spatial distance, and $\omega_{mn}$ denotes the "dielectric constant". $\omega_{mn}$ can be time dependent, so $T_{mn}$ can be time dependent, and $F_{mn}$ can also be time dependent. Equation (1) is similar to the gravitational equation [1,2,3,4,5,6], but instead of spatial distance, bilateral distance is used.

Consider a country in position $T$ with a GDP $G$. We define the "trade strength" at the center of the coordinates due to the GDP $G$ as

$$g = K\frac{G^\delta}{T^\beta} \quad (2)$$

The $g$ variable simulates gravity's acceleration at the coordinates' center. Equation (2) is obtained by dividing Eq. (1) by $G_n^\delta$, i.e., $g_m = F_{mn}/G_n^\delta$.

We assume that $G$ does not change significantly over time [27]. We will derive the distribution of $g$ and $f$(the rate of the $g$),

$$f = \frac{dg}{dt} = -K\beta\frac{G^\delta}{T^{\beta+1}}\frac{dT}{dt} - K\beta\frac{G^\delta}{T^{\beta+1}}v \quad (3)$$

where $v = dT/dt$. Assume that the probability of the distance and "speed" of the trade strength between $T$ to $T + dT$ and $v$ to $v + dv$ is $\tau(T, v)dTdv$. The probability distribution of trade strength $g$ in the range of $g_0 \le g \le g_0 + dg_0$ and its rate $f$ in the range of $f_0 \le f \le f_0 + df_0$ can be expressed as

$$w(g_0, f_0, G)dg_0 df_0 = \frac{1}{A}\int\int_{g_0 \le g \le g_0 + dg_0, f_0 \le f \le f_0 + df_0} \tau(T, v)dTdv \quad (4)$$

where $A$ is the total "phase" area of all countries. Let us employ the Dirichlet factor, which has a value of one if the preceding constraint is satisfied and zero if it is not, i.e. [26,28]

$$\frac{1}{\pi^2}\int_{-\infty}^{\infty}\int_{-\infty}^{\infty} e^{i\rho(g-g_0)} e^{i\sigma(f-f_0)} \frac{\sin\left(\frac{1}{2}\rho g\right)}{\rho} \frac{\sin\left(\frac{1}{2}\sigma f\right)}{\sigma} d\rho d\sigma = \begin{cases} 1 \quad if \quad g_0 \le g \le g_0 + +dg_0 \\ \quad\quad and\ f_0 \le f \le f_0 + df_0 \\ 0 \quad if \quad others \end{cases} \tag{5}$$

Using Eq. (5), we can write Eq. (4) as

$$w(g_0, f_0, G) dg_0 df_0 = \frac{dg_0 df_0}{4\pi^2} \int_{-\infty}^{\infty}\int_{-\infty}^{\infty} e^{-i\rho g_0} e^{-i\sigma f_0} B(\rho, \sigma, G) d\rho d\sigma \tag{6}$$

where

$$B(\rho, \sigma, G) = \frac{1}{A}\int_{T=0}^{\infty}\int_{v=-\infty}^{\infty} e^{i\rho g} e^{i\sigma f} \tau(T, v) dT dv \tag{7}$$

We have taken the integral limits for $T$ to be from zero to infinity, while the integral limits for $v$ are from negative infinity to positive infinity.

Assume that $\tau$ fulfills the general equation $\tau = (q/T^\theta)e^{-p^2v^2}$, where $p$, $q$, and $\theta$ are all positive constants. It is obvious that the greater the distance and "speed," the lower the probability. With this assumption, Eq. (7) can be written as

$$B(\rho, \sigma, G) = \frac{q}{A}\int_0^{\infty} \frac{1}{T^\theta} e^{i\rho g} dT \int_{-\infty}^{\infty} e^{-i\sigma K\beta \frac{G^\delta}{T^{\beta+1}} v} e^{-p^2v^2} dv$$

$$= \frac{q\sqrt{\pi}}{Ap}\int_0^{\infty} \frac{1}{T^\theta} e^{i\rho K \frac{G^\delta}{T^\beta}} \exp\left[-\frac{K^2\beta^2 G^{2\delta}\sigma^2}{4p^2 T^{2\beta+2}}\right] dT \tag{8}$$

after using the identity as described in [26]. We can complete the integration in Eq. (8) easily if we know the value of several parameters, such as $\beta$. At present we approximate $\beta \approx 3/2$ because it is easier to find analytical solutions with this parameter. This figure is not significantly different from the estimate in **Supplementary 1** ($\beta \approx 1.7$). Several reports about the value of the $\beta$ are 0.72 [2], 0.942 [7] and 0.924 [9]. Gul and Yasin [10] have reported the various values of β for Pakistan's trade with EU ($\beta = 1$), Pakistan's trade vs ASEAN ($\beta = 0.81$), Pakistan's trade vs SAAR-ECO ($\beta = 0.35$), Pakistan's trade vs Middle East ($\beta = 7.99$), Pakistan's trade vs Far East ($\beta = 0.77$), and Pakistan's trade vs NAFTA and Latin America ($\beta = 1.93$). Other reported parameters are β = 0.956 [21], β = 1 [22], β = 0.705 [23], and β = 1.281 [24]. By substituting the proposed $\beta \approx 3/2$, we get

$$B(\rho,\sigma,G) = \frac{q\sqrt{\pi}}{Ap}\int_0^\infty \frac{1}{T^\theta} e^{i\rho K\frac{G^\delta}{T^{3/2}}} \exp\left[-\frac{9K^2G^{2\delta}\sigma^2}{16p^2T^5}\right] dT \qquad (9)$$

Let us further define $z^{10/3} = 9K^2G^{2\delta}\sigma^2/16p^2T^5$ so that Eq. (6) becomes

$$w(g,f,G) = \frac{1}{4\pi^2}\frac{2q\sqrt{\pi}}{3Ap}\left(\frac{9K^2G^{2\delta}}{16p^2}\right)^{(1-\theta)/5}$$

$$\times \int_{-\infty}^{\infty} e^{-i\sigma f}\sigma^{2(1-\theta)/5}\int_0^\infty z^{(2\theta-5)/3}e^{-z^{10/3}}\int_{-\infty}^{\infty} e^{i\rho\left[\left(\frac{16p^2}{9K^2G^{2\delta}\sigma^2}\right)^{3/10}KG^\delta z - g\right]} d\rho\, dz d\sigma \qquad (10)$$

In the distribution function, we replaced $g_0$ with $g$ and $f_0$ with $f$. Following that, we employ the Dirac delta function's definition and its property, $\delta(x-x_0) = (1/2\pi)\int_{-\infty}^{\infty} e^{i\omega(x-x_0)}d\omega$ and $\delta\big(k(x-x_0)\big) = (1/|k|)\delta(x-x_0)$ to yield

$$w(g,f,G) = \frac{q}{3Ap\sqrt{\pi}}\left(\frac{9K^2G^{2\delta}}{16p^2}\right)^{(1-\theta)/5}\frac{1}{KG^\delta}\left(\frac{9K^2G^{2\delta}}{16p^2}\right)^{3/10}\times$$

$$\int_{-\infty}^{\infty} e^{-i\sigma f}\sigma^{2(1-\theta)/5}\int_0^\infty z^{(2\theta-5)/3}e^{-z^{10/3}}\sigma^{3/5}\delta\big(z-\omega g\sigma^{3/5}\big)\, dz d\sigma \qquad (11)$$

where we have defined

$$\omega = \frac{1}{KG^\delta}\left(\frac{9K^2G^{2\delta}}{16p^2}\right)^{3/10} \propto G^{-4\delta/10} \qquad (12)$$

We first integrate on the variable $z$ and pursuing a transformation $y^2 = \omega^{10/3}g^{10/3}\sigma^2$ to yield

$$w(g,f,G) = \frac{q}{3Ap}\left(\frac{9K^2G^{2\delta}}{16p^2}\right)^{\frac{1-\theta}{5}}\frac{1}{KG^\delta}\left(\frac{9K^2G^{2\delta}}{16p^2}\right)^{\frac{3}{10}}\omega^{\frac{2\theta-10}{3}}g^{\frac{2\theta-10}{3}}\exp\left[-\frac{f^2}{4\omega^{10/3}g^{10/3}}\right] \qquad (13)$$

For example, suppose $\theta = 1$ so that $\tau = (q/T)e^{-p^2v^2}$. We don't know what the exact value of $\theta$ is. In their report, Chandrasekhar and Neumann used $\theta = 0$ [26], which means that the distribution depends only on velocity but does not depend on distance at all. We do, however, believe that distance plays a role in determining trading fluctuations. The greater the distance, the smaller the fluctuations. With this hypothesis, we get

$$w(g,f,G) = \frac{q}{3ApKG^\delta}\left(\frac{9K^2G^{2\delta}}{16p^2}\right)^{\frac{3}{10}}\omega^{-\frac{8}{3}}g^{-\frac{8}{3}}\exp\left[-\frac{f^2}{4\omega^{\frac{10}{3}}g^{\frac{10}{3}}}\right] \propto \frac{G^{\frac{2\delta}{3}}}{g^{\frac{8}{3}}}\exp\left[-\frac{\kappa G^{\frac{4\delta}{3}}}{4g^{\frac{10}{3}}}f^2\right] \qquad (14)$$

after substituting Eq. (12).

The distribution of the trade strength as expressed by Eq. (14) is contributed by one country only. If the GDP distribution function is expressed as $f(G)$, the average distribution, $\overline{w(g,f,G)}$, becomes

$$w(g,f) = \overline{w(g,f,G)} \propto \frac{1}{g^{\frac{8}{3}}} \int_0^\infty G^{2\delta/3} \exp\left[-\frac{\kappa G^{\frac{4\delta}{3}}}{4g^{\frac{10}{3}}} f^2\right] f(G) dG \qquad (15)$$

Equation (15) can be solved if we know the distribution function and the parameter $\delta$ ind the GDP. The value of these parameters will be estimated using international trade data from several countries [29].

We will estimate the distribution function, $f(G)$, using World Bank GDP data for all countries [20]. We use the GDP data of year 2005, 2010, 2015, and 2020. Given that GDP is always positive, the normal log distribution is one of candidate of the distribution functions. In **Supplementary 2**, we show that the GDP distribution fits quite well with a lognormal distribution.

To make the integral in Eq. (15) easier to calculate, let us expand the exponential in the series so that we can write

$$w(g,f) \propto \frac{1}{g^{8/3}} \sum_{j=0}^{\infty} (-1)^j \frac{1}{j!} \left(\frac{\kappa}{4g^{\frac{10}{3}}} f^2\right)^j \int_0^\infty G^{4j\delta/3+2\delta/3} f(G) dG \qquad (16)$$

Using the log normal distribution's properties and some approximations (**Supplementary 3**), we finally get the distribution function$w(g,f)$ as

$$w(g,f) \approx \frac{B}{g^{\frac{8}{3}}} \exp\left(-\frac{(f/\kappa')^2}{g^{\frac{10}{3}}}\right) \qquad (17)$$

where $\kappa' = 2/\sqrt{\kappa e^{8\mu/9}}$ and $B$ is a constant.

## III. RESULTS AND DISCUSSION

Figure 1 depicts $w(g,f)$ as a function of $g$ and $f$:(a): surface plot and (b) contour plot. Figure 1(c) depicts the $w$ on $g$ dependency curve for various $f/\kappa'$. For a given $f$, it appears from Eq. (17) that $w$ has a peak at a value $g$ that meets the condition $\partial w(g,f)/\partial g = 0$. Using Eq. (17), the peak occurred at $g$, which satisfies

$$g \approx \left(\frac{f}{\kappa'}\right)^{3/5} \tag{18}$$

Equation (17) clearly demonstrates that if $f \to 0$, $w$ fulfills the scaling relationship

$$w \propto g^{-8/3} \tag{19}$$

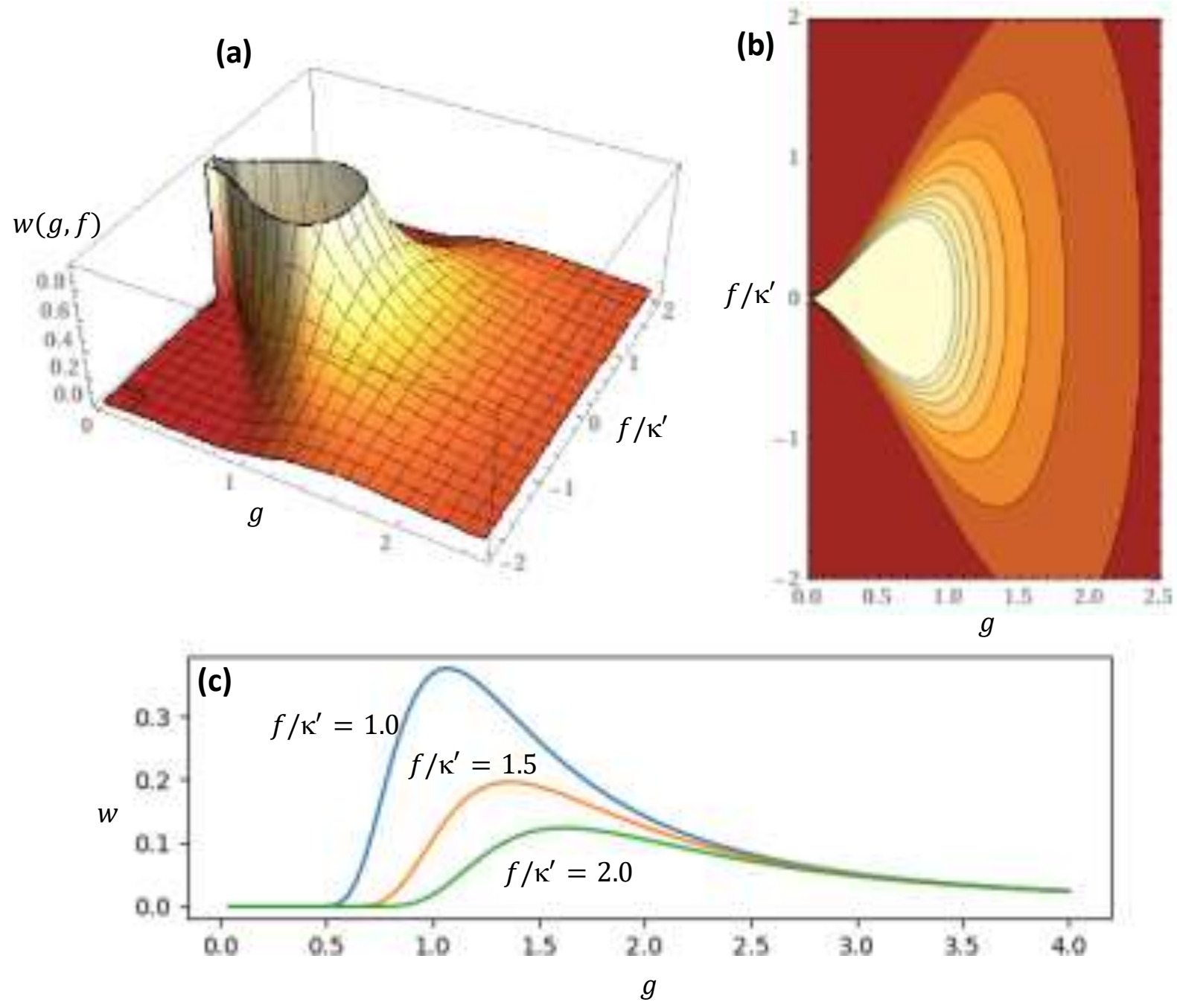

FIG. 1. (a) surface plot $w(g,f)$ dan contour plot $w(g,f)$. (c) Dependence of $w$ on $g$ on various selected values of $f/\kappa'$.

If we define a variable $x$ that satisfies $x^2 = 1/g^{10/3}$, we can write Eq. (1) as

$$w(x,f) = w(g,f)\left|\frac{dg}{dx}\right| \approx \frac{5}{3}Bx^{16/9}\exp\left(-\frac{1}{\kappa'^2}f^2x^2\right) \tag{20}$$

So, for a given $x$, $w(x,f)$ satisfies the normal distribution function with respect to $f$. However, for a given $f$, $w(x,f)$ changes with $x$ to resemble the Maxwell-Boltzmann distribution equation for speed in three dimensions. In the Maxwell-Boltzmann distribution, we get $f(v) \propto v^2 \exp(-mv^2/2kT)$. The power for speed is two, while for our case, the power for $x$ is 16/9 = 1.78.

Let us now examine $w$ dependence on each variable $g$ or $f$. We obtain from Eq. (17)

$$w(g) = \int_{-\infty}^{\infty} w(g,f)df = \frac{B\kappa'\sqrt{\pi}}{g} \quad (21)$$

$$w(f) = \int_{0}^{\infty} w(g,f)dg = \frac{3B\kappa'\sqrt{\pi}}{10f} \quad (22)$$

The $w(g)$ curve as a function of $1/g$ is depicted in Fig. 2. The following are the steps for calculating $g$ and $w(g)$. The value of $g$ for each country is determined by Eq. (2), which can be written as $g = F/G^{\delta}$, where $F$ is the total trade volume (export + import) of a country with all other countries in a specific year, $G$ is the country's GDP in the same year, and $\delta = 0.665$ (see **Supplementary 1**). For total trade, we use data from [29], and for GDP, we use data from [20]. The distribution function is then determined by dividing $g$ over several ranges. We used data from 163 countries in the years preceding the Covid-19 pandemic, specifically 2008, 2010, 2012, 2014, 2016, and 2018, and the results are shown in Fig. 2.

The data is fitted with a linear function, and the intercept is set to zero, as shown by the straight line in Fig. 2. The fitting results appear to be quite good, indicating that the distribution function obtained is justified. It's fascinating that the slopes of all fitting lines are nearly identical, namely 0.2. This result is consistent with Eq. (21) in which the slope is constant, namely $B\kappa'\sqrt{\pi}$. We can also deduce from this result that $B\kappa' \cong 0.113$.

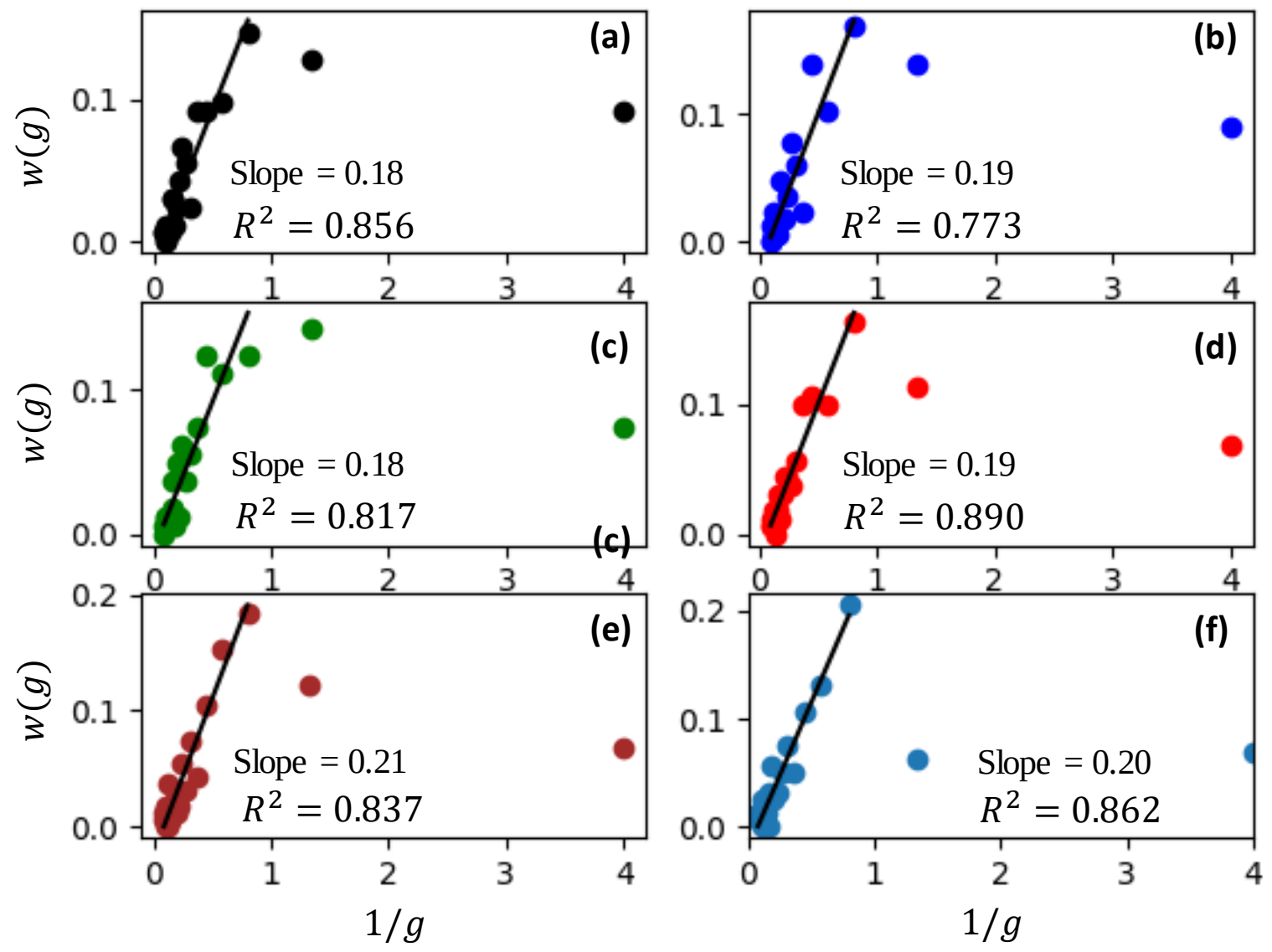

FIG. 2. $w$ dependence on $1/g$ in different years: (a) 2008, (b) 2010, (c) 2012, (d) 2014, (e) 2016, and (f) 2018. The symbols are data from [20], and the lines are the results of linear fitting excluding two data points at $1/g$ large.

To test the dependence of $w$ on $f$ as shown in Eq. (22), we take $f$ proportional to each country's GDP growth (see **Supplementary 4**), i.e.

$$f_j \propto \frac{1}{G_j}\frac{dG_j}{dt} \tag{23}$$

Figure 3 depicts the dependence of $w$ on $f$ as expressed in Eq. (22). We use data from the following years: (a) 2008, (b) 2010, (c) 2012, (d) 2014, (e) 2016, and (f) 2018. To avoid the effects of the Covid-19 pandemic on GDP and growth, we do not use data after 2018. For high $f$, it appears that $w$ changes linearly with $1/f$, which is consistent with Eq. (22). Linear fitting of data with large $f$ produces nearly the same slope, namely $\approx 0.2$ across all years. This result is consistent with the constant scaling coefficient in Eq. (22).

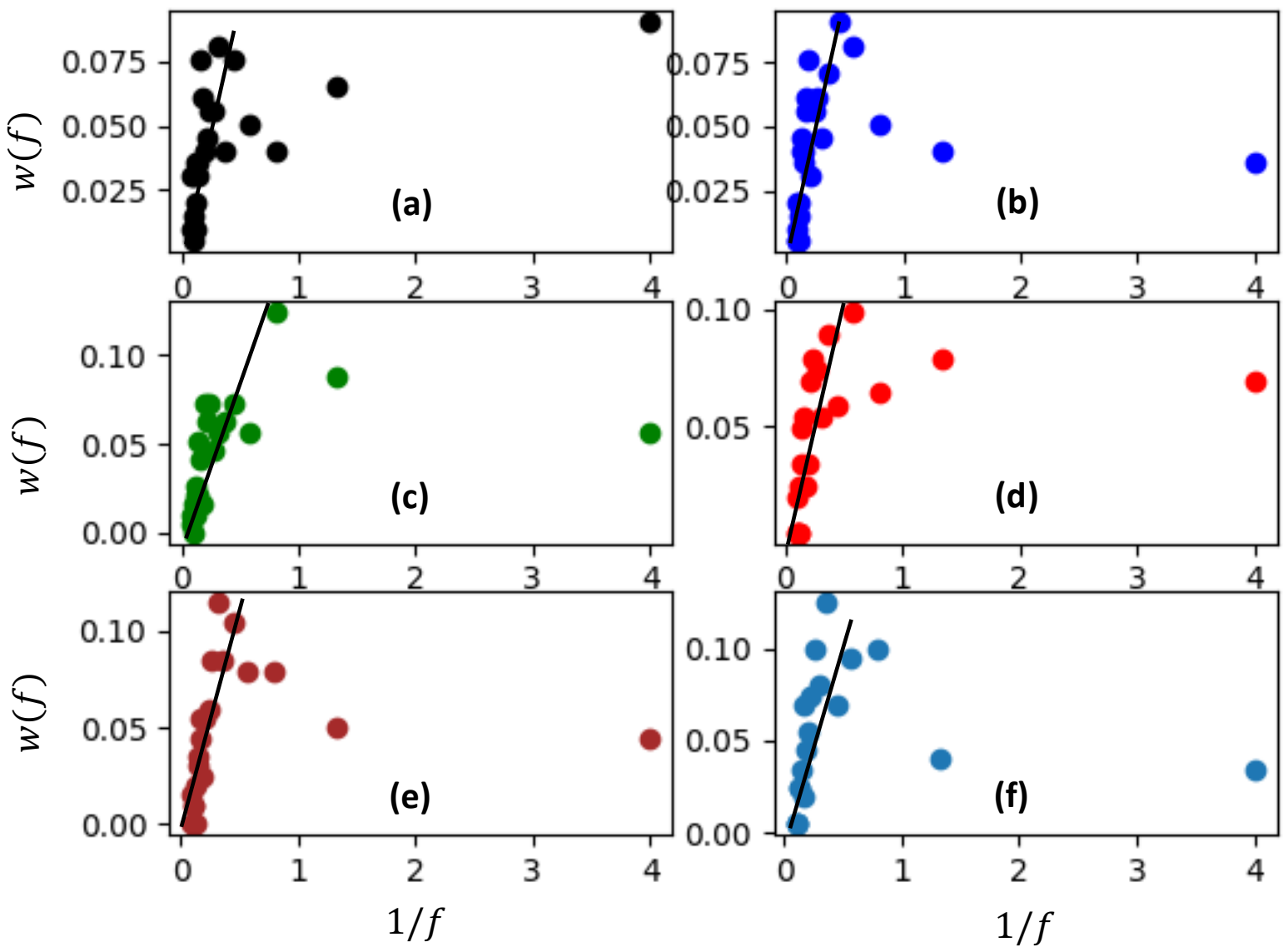

FIG. 3. The dependence of $w$ on $1/f$ in various years: (a) 2008, (b) 2010, (c) 2012, (d) 2014, (e) 2016, and (f) 2018. Symbols are data obtained from [20] and lines are the results of linear fitting excluding any two data at large $1/f$. The process for estimating $f$ is described in **Supplementary 4**.

We haven't figured out why the data for countries with small $f$ and $g$ (large $1/f$ and $1/g$) deviate far enough from the fitting curve. There are two data points that deviate from the fitting curve in each of Fig. 2 and 3 which are contributed by countries with low GDP. The number of countries covered by the data is approximately 20% of the total countries used in the calculation, but their total GDP is only approximately 4% of the total GDP of all countries used in the calculation. As a result, ignoring the two data points during the fitting process is acceptable.

In relation to Eqs. (21) and (22), the slopes in Eq. (22) (Fig. 3) appear to be 3/10 of the slopes in Eq. (21) (Fig. 2). However, the fitting results show that the slopes of the fitting lines in Figs. 2 and 3 are nearly identical, i.e., around 0.2. The slopes cannot be directly compared because we did not use the proper units for $g$ and $f$ when creating the curves in Figs. 2 and 3. (we still use arbitrary units). What we want to do now is to test Eqs. (21) and (22) to see if $w$ changes linearly with $1/g$ and clinearly with $1/f$. We also want to show that the slope of the fitting curve $w$ to $1/g$ is always the same in all years and that the slope of the fitting curve $w$ to $1/f$ is always the same in all years. We have demonstrated that both properties are met, implying that the distribution function in Eq. (17) can be accepted.

## IV. CONCLUSION

We conclude that the coulomb force model can explain trade flow using a simpler equation than the gravity equation. All non-metric parameters that influence trade flow are incorporated into the "dielectric constant". The trade strength distribution function developed here for the first time can explain data from several countries over a decade. These findings are expected to serve as a guide for improving the economic welfare of countries with low GDP per capita.

(*)Email: mikrajuddin@gmail.com

# Supplementaries:

## Coulomb-like Model for International Trade Flow and Derivation of Distribution Function for Trade Flow Strength

### Supplementary 1

**Derivation of Coulomb Equation**

Let's start by deriving the coulomb equation to describe trading volume between countries. Figure s1 is an illustration of a country's export and import flow. Suppose there are $N$ countries. In year $t$, the country $m$ exports goods to several other countries. Suppose the export volume to a country $q$ is $E_{mq}(t)$. The total exports of country m to all other countries in year $t$ can be expressed as

$$E_m(t) = \sum_{q=1}^{N} E_{mq}(t) \quad \text{(s1)}$$

where $E_{mm}(t) = 0$. We suppose the country $m$ imports from the country $r$ in the $t$-th year with the value $I_{mr}(t)$. The total imports of country $m$ from all other countries in year $t$ become

$$I_m(t) = \sum_{r=1}^{N} I_{mr}(t) \quad \text{(s2)}$$

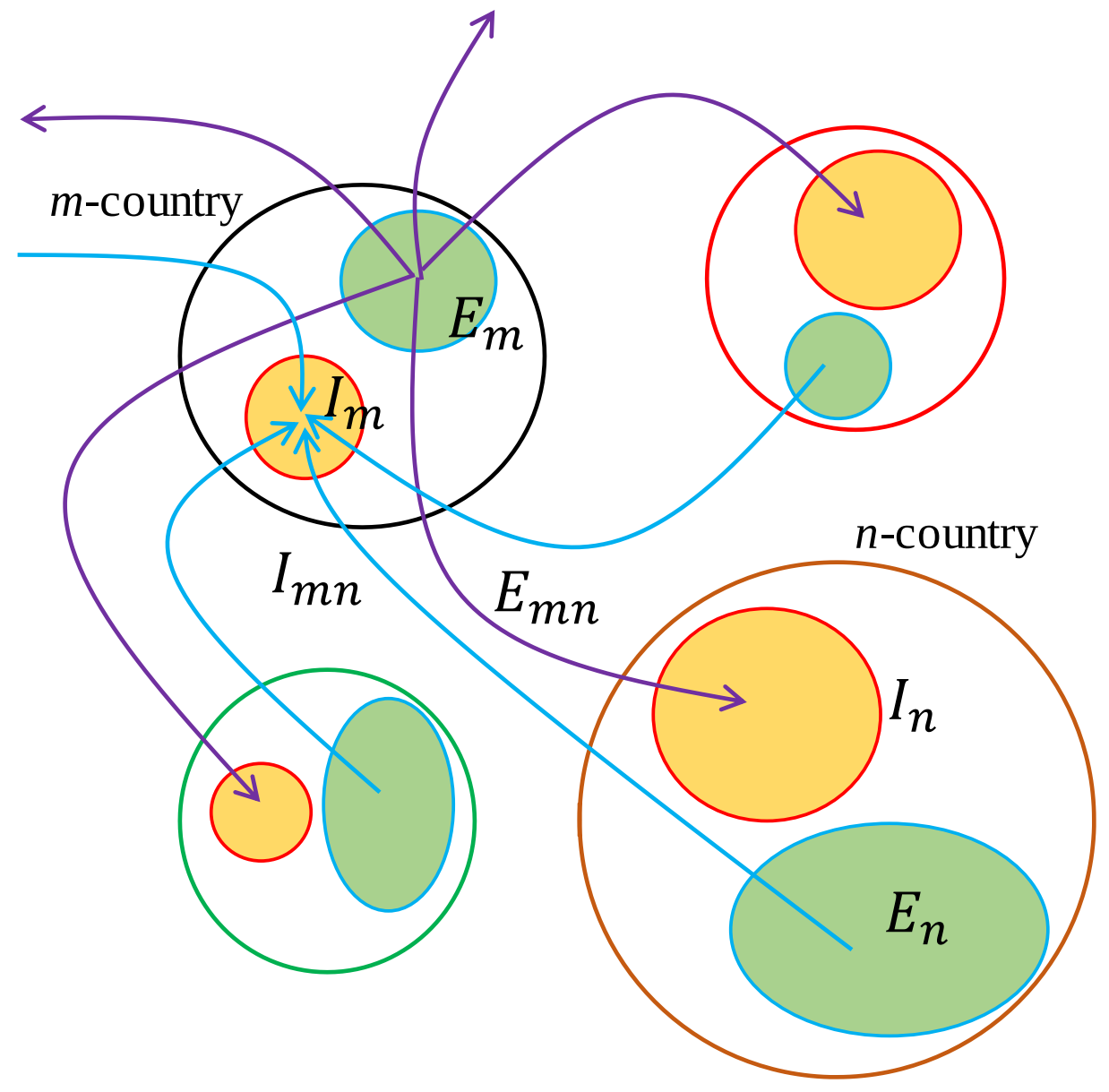

FIG. s1. Illustration of the flow of exports and imports between countries.

Let us consider the flow of trade between country $m$ and country $n$. We hypothesize that the value of exports from country $m$ to country $n$ in year $t$ depends on the total volume of exports in year $t$ owned by country $m$ and the total volume of imports owned by country $n$. We postulate that the value of the export depends on the product of $(E_m(t)I_n(t))^\alpha$ where $\alpha$ is an unknown parameter. The same hypothesis is that the import value of country $m$ from country $n$ in year $t$ depends on the volume of imports of country $m$ and the volume of exports of country $n$, in the form $(I_m(t)E_n(t))^\alpha$. Thus, we hypothesize that the value of trade between country $m$ and country $n$ in year $t$ satisfies the equation

$$F_{mn}(t) = F_{m\to n}(t) \propto [a(E_m(t)I_n(t))^\alpha + b(I_m(t)E_n(t))^\alpha] \quad \text{(s3)}$$

where $a$ and $b$ are constants that are generally independent of time.

The trade value also depends on the distance between the two countries, $R_{mn}$, and the condition of the relationship between the two countries, $\omega_{mn}(t)$. Thus, we can write the general equation for the trade value of country $m$ and country $n$ in year $t$ satisfying the general equation

$$F_{mn}(t) = \frac{1}{\omega_{mn}(t)} \frac{[a(E_m(t)I_n(t))^\alpha + b(I_m(t)E_n(t))^\alpha]}{R_{mn}^\beta} \quad \text{(s4)}$$

where $\omega_{mn}$ depends on the restrictions on the movement of goods across international borders, and $\beta$ is unknown parameter. For a free trade, the value of $\omega_{mn}$ is small. From Eq. (s4) we can write the reverse trade, i.e. from country $n$ to country $m$, i.e.,

$$F_{nm}(t) = F_{n\to m}(t) = \frac{1}{\omega_{nm}(t)} \frac{[a(E_n(t)I_m(t))^\alpha + b(I_n(t)E_m(t))^\alpha]}{R_{nm}^\beta} \quad \text{(s5)}$$

Since $F_{nm}(t) = F_{mn}(t)$, $\omega_{nm}(t) = \omega_{mn}(t)$, and $R_{nm} = R_{mn}$, we get

$$a(E_m(t)I_n(t))^\alpha + b(I_m(t)E_n(t))^\alpha = a(E_n(t)I_m(t))^\alpha + b(I_n(t)E_m(t))^\alpha \quad \text{(s6)}$$

This equation is automatically satisfied if $a = b$ and Eq. (s6) can be written as

$$F_{mn}(t) = \frac{1}{\omega_{mn}(t)} \frac{(E_m(t)I_n(t))^\alpha + (I_m(t)E_n(t))^\alpha}{R_{mn}^\beta} \quad \text{(s7)}$$

where $\omega_{mn}(t)$ here has absorbed the constant $a$ into it. Let us take the logarithm on both sides of Eq. (s7) to get

$$\ln F_{mn}(t) = \ln[(E_m(t)I_n(t))^\alpha + (I_m(t)E_n(t))^\alpha] - \beta \ln R_{mn} - \ln \omega_{mn}(t) \qquad \text{(s8)}$$

Next we intend to determine the value of the parameters $\alpha$ and $\beta$. We do not use the panel data to find parameters $\alpha$ and $\beta$, but use a method commonly used in physics. To determine the parameter $\alpha$, we do the following steps:

**(1)** We choose the bilateral trades of two specific countries over several years. For a given country pair, $R_{mn}$ is fixed for all time. If there is no significant change in the trade agreement or political issues between the two countries, $\omega_{mn}$ between the two countries can be considered constant. Thus, the change in trade value at various times is solely determined by the first term on the right side of Eq. (9).

**(2)** For each country pair, we choose the value of $\alpha$ such that the curve of $\ln Trade_{mn}$ changes linearly against $\ln[(E_m(t)I_n(t))^\alpha + (I_m(t)E_n(t))^\alpha]$ with a slope equal to one. We do this for as many pairs of countries and check whether the obtained $\alpha$ are concentrated around a certain value. If so, we can assume that the proposed equation is acceptable and the mean value of $\alpha$ is considered valid.

To determine the parameter $\beta$, we perform the following steps.

**(1)** From the two pairs of countries that carry out bilateral trade, Eq. (s7) can be written as follows,

$$\frac{F_{mn}(t)}{F_{pq}(t)} = \frac{\omega_{pq}(t)}{\omega_{mn}(t)} \frac{(E_m(t)I_n(t))^\alpha + (I_m(t)E_n(t))^\alpha}{\left(E_p(t)I_q(t)\right)^\alpha + \left(I_p(t)E_q(t)\right)^\alpha} \left(\frac{R_{pq}}{R_{mn}}\right)^\beta \qquad \text{(s9)}$$

**(2)** We assume that countries that form certain economic agreements such as EC, ASEAN, or NAFTA have similar trade rules between member countries. Thus, the dielectric constant between the member can be considered the same and Eq. (s9) can be approximated as

$$\frac{F_{mn}(t)}{F_{pq}(t)} \approx \frac{(E_m(t)I_n(t))^\alpha + (I_m(t)E_n(t))^\alpha}{\left(E_p(t)I_q(t)\right)^\alpha + \left(I_p(t)E_q(t)\right)^\alpha} \left(\frac{R_{pq}}{R_{mn}}\right)^\beta$$

or

$$\ln\left(\frac{F_{mn}(t)}{F_{pq}(t)}\right) \approx \ln\left[\frac{\left(E_m(t)I_n(t)\right)^{\alpha}+\left(I_m(t)E_n(t)\right)^{\alpha}}{\left(E_p(t)I_q(t)\right)^{\alpha}+\left(I_p(t)E_q(t)\right)^{\alpha}}\right] + \beta\ln\left(\frac{R_{pq}}{R_{mn}}\right) \tag{s10}$$

**(3)** Next we draw the curve $\ln\left(\frac{F_{mn}(t)}{F_{pq}(t)}\right)$ against $\ln\left[\frac{\left(E_m(t)I_n(t)\right)^{\alpha}+\left(I_m(t)E_n(t)\right)^{\alpha}}{\left(E_p(t)I_q(t)\right)^{\alpha}+\left(I_p(t)E_q(t)\right)^{\alpha}}\right]$. We fit linearly the obtained data and the intersection of the curve with the axis $\ln\left(\frac{F_{mn}(t)}{F_{pq}(t)}\right)$ is the value of $\beta\ln\left(R_{pq}/R_{mn}\right)$. By entering the ratio of the distance between the two pairs of countries (capital cities), the value of $\beta$ can be estimated.

### Data Sources

We use data from the World Integrated Trade Solution [3] for several groups of countries with large economies in ASEAN (Indonesia, Singapore, Malaysia, Vietnam, Thailand, Philippines, and Vietnam), G7 (USA, Germany, UK, France, Italy, Japan, and Canada), NAFTA (USA, Canada, and Mexico), Mercosur (Brazil, Argentina, Uruguay, and Paraguay), and East Asia (China, Japan, South Korea, and Hongkong).

### Estimation of Parameter α

Figure s2 is an example of the dependence of $y = \ln F_{mn}(t)$ as a function of $x = \ln[(E_m(t)I_n(t))^{\alpha} + (I_m(t)E_n(t))^{\alpha}]$ a number of pairs of ASEAN countries. We used data from 2009-2019. The parameter $\alpha$ has been chosen such that the data changes linearly and the fitting curve has a slope close to one (ideally, the slope is one). It appears that to obtain a fitting curve with a slope of unity, the parameter $\alpha$ varies for different pairs of countries. However, the value of the obtained parameter $\alpha$ fluctuates around 0.5. Figure s3 is an example of a similar curve for NAFTA member countries. It is also seen here that the obtained parameter $\alpha$ is around 0.5.

We have calculated the $\alpha$ parameters obtained from the pair of countries analyzed (ASEAN, G7, NAFTA, Mercusor, and East Asia). The value of the $\alpha$ parameter varies from 0.24 to 0.75, with the average $\mu = 0.47$. These results are close to those reported by previous researchers such as $\alpha_1 = \alpha_2 = 0.728$ [4], $\alpha_1 = 0.80$ and $\alpha_2 = 0.65$ [5], $\alpha_1 = 0.922$ and $\alpha_2 =$ 0.930 [6], $\alpha_1 = \alpha_2 = 0.67 - 0.91$ [7]. Gul and Yasin [8] have reported the following parameters for Pakistan's trade with other countries: Pakistan vs all countries ($\alpha_1 = \alpha_2 = 0.89$), Pakistan vs EU ($\alpha_1 = \alpha_2 = 0.97$), Pakistan vs ASEAN ($\alpha_1 = \alpha_2 = 0.65,$), Pakistan vs SAAR-ECO ($\alpha_1 = \alpha_2 = 0.61,$), Pakistan vs Middle East ($\alpha_1 = \alpha_2 = 0.92,$), Pakistan vs Far East ($\alpha_1 = \alpha_2 = 0.66$), and Pakistan vs NAFTA and Latin America ($\alpha_1 = \alpha_2 = 1.65$). In the next analysis, we will use $\alpha = \alpha_1 \approx \alpha_2 \approx 0.5$.

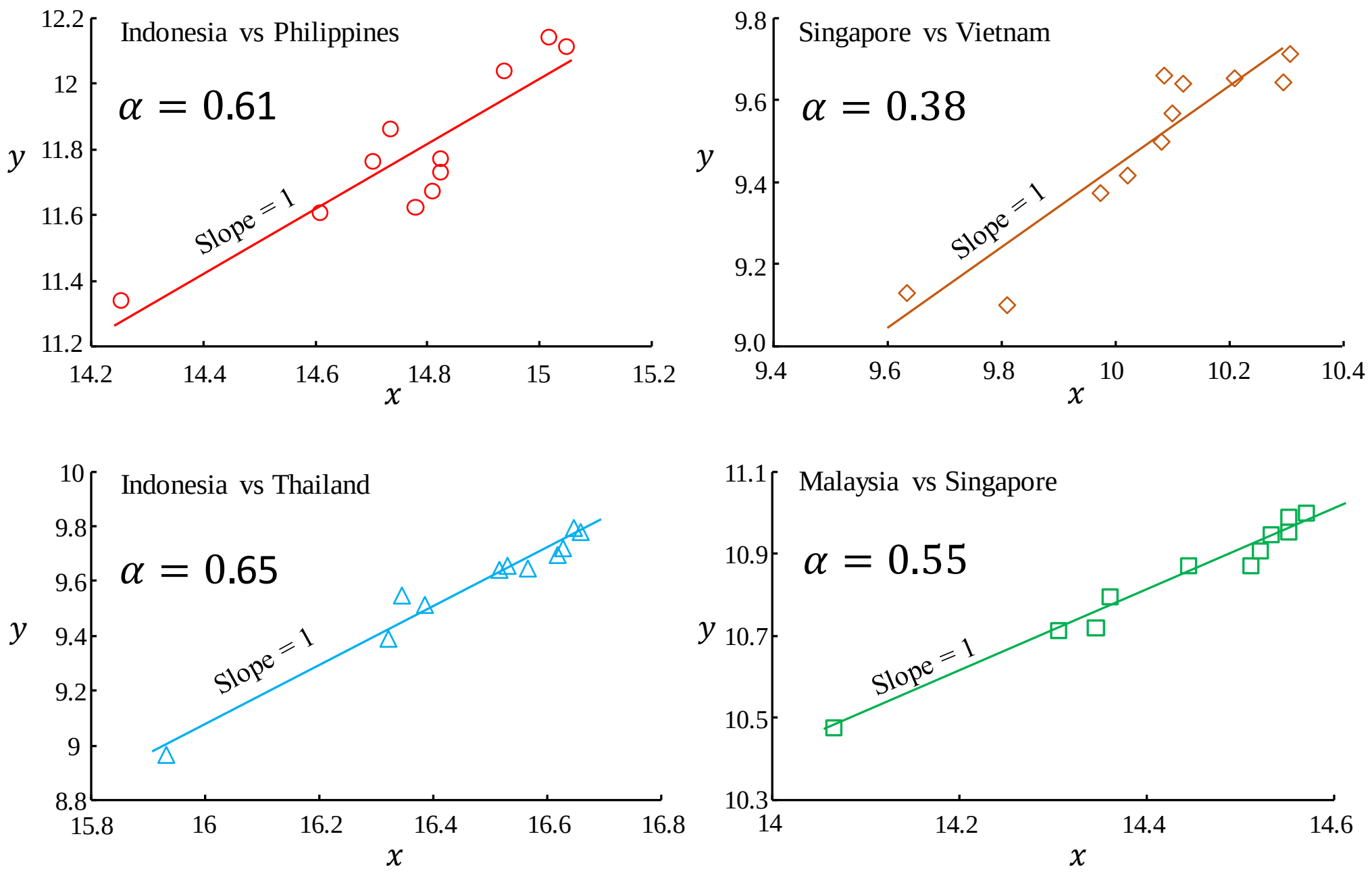

FIG. s2. The relationship between $x$ and $y$ of a number of ASEAN member countries between 2009-2019. See text for definitions of $x$ and $y$. Symbols are data from [3] and lines are the result of linear fitting.

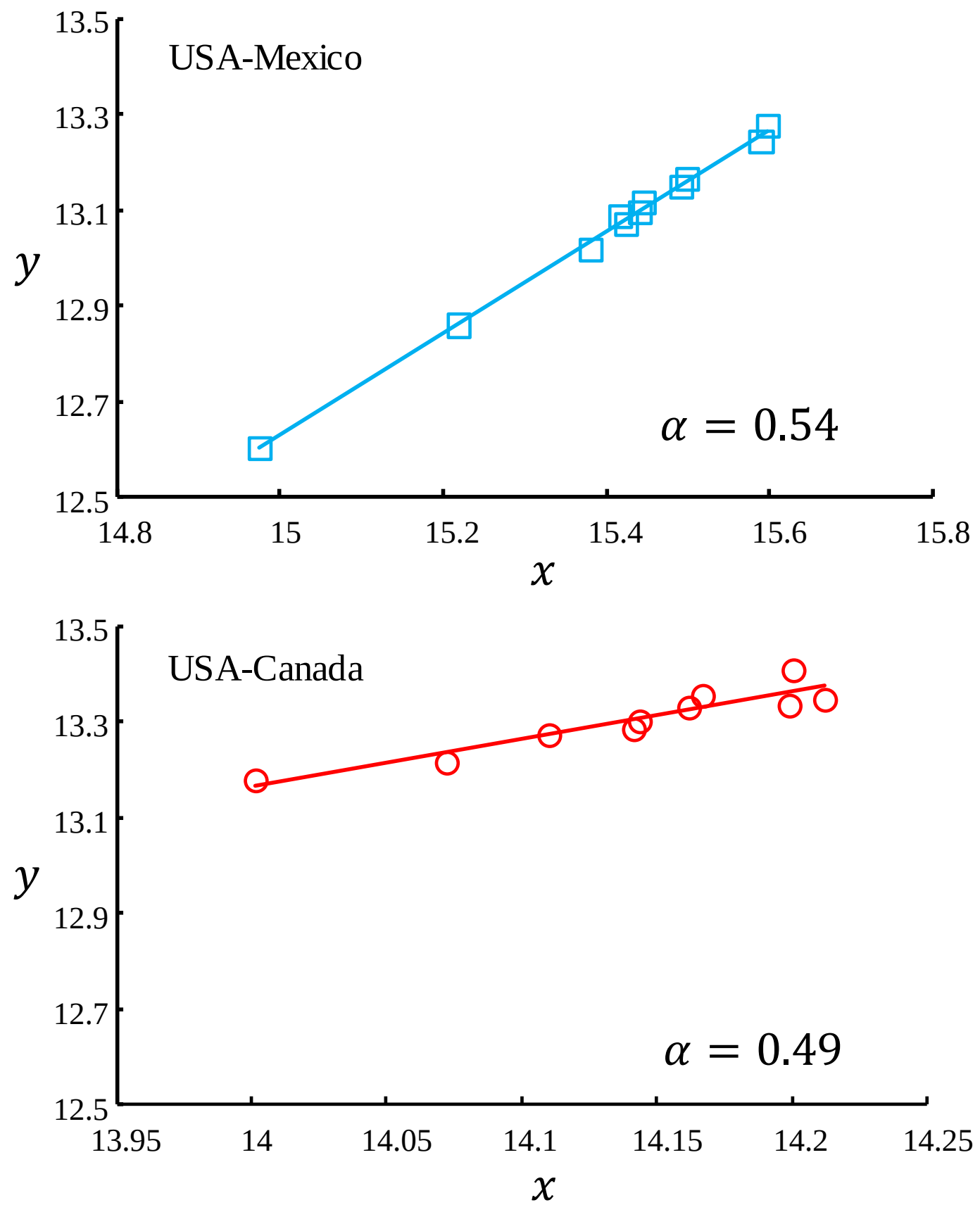

FIG. s3. The relationship between $x$ and $y$ of a number of NAFTA member countries between 2009-2019. See text for definitions of $x$ and $y$. Symbols are data from [3] and lines are the result of linear fitting.

**Table A1** shows a complete list of the $\alpha$ parameters obtained from the countries analyzed (ASEAN, G7, NAFTA, Mercusor, and East Asia). It can be seen from the table that the value of the $\alpha$ parameter varies from 0.24 to 0.75.

### Estimation of Parameter $\beta$

Next we will estimate the parameter $\beta$. This parameter is not very accurate to be estimated because it is very sensitive to the influence of other countries besides the two pairs of countries calculated. To estimate this parameter, we take pairs of countries where the trade of each pair is almost not significantly affected by countries other than them. We need three "isolated" countries that enter into trade agreements, but trade between one pair with another pair does not affect each other. One region approaching this condition is NAFTA consisting of three countries: the USA, Canada and Mexico. The three countries are in a diametrical position where the USA is in the middle, Canada is in the north, and Mexico is in the south. Trade between USA-Canada and USA-Mexico is very large, but trade between Canada-Mexico is very small. Thus, the Canada-Mexico interaction is negligible. Therefore we can use the USA-Canada and USA-Mexico trade data to estimate the $\beta$ parameter.

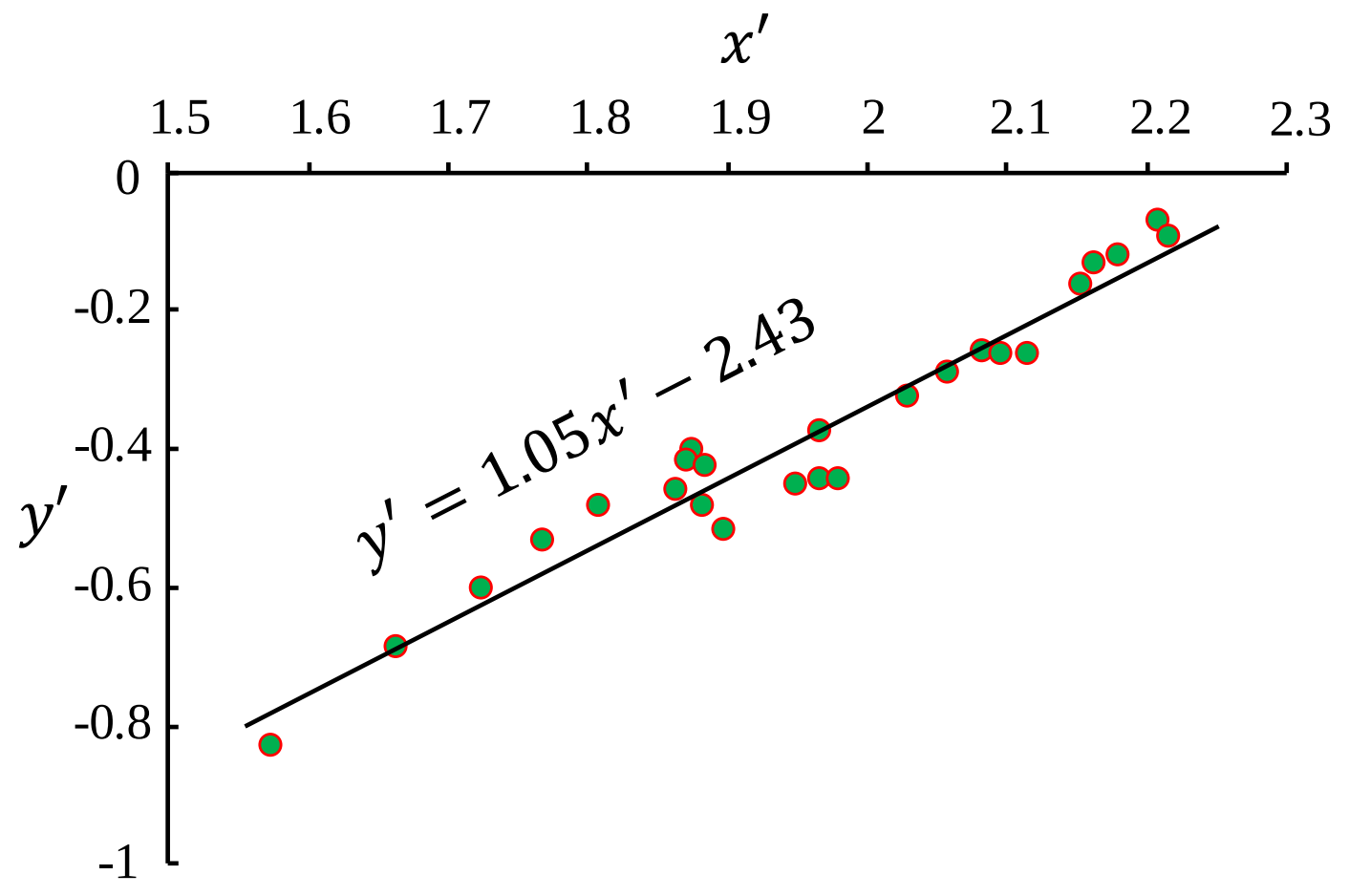

FIG. s4. USA-Canada and USA-Mexico trade ratios. See text for descriptions of $x'$ and $y'$. Symbols are data sourced from [3] and line is the result of linear fitting.

Figure s4 is the curve $y' = \ln\left(\frac{F_{mn}(t)}{F_{mq}(t)}\right)$ against $x' = \ln\left[\frac{\left(E_m(t)I_n(t)\right)^{\alpha_{mn}} + \left(I_m(t)E_n(t)\right)^{\alpha_{mn}}}{\left(E_m(t)I_q(t)\right)^{\alpha_{mq}} + \left(I_m(t)E_q(t)\right)^{\alpha_{mq}}}\right]$

where $m$ = USA, $n$ = Canada, and $q$ = Mexico. Based on the data in the **Table A1**, we have $\alpha_{mn} = 0.47$ and $\alpha_{mq} = 0.55$. We fit the data obtained with a linear curve and obtained the fitting equation $y' = 1.05x' - 2.43$. What is most important for us is the pivot point of the curve with the vertical axis, i.e.

$$\beta \ln\left(\frac{R_{mq}}{R_{mn}}\right) = -2.43 \qquad \text{(s11)}$$

We used $R_{mn}$ as the Washington DC-Ottawa distance and $R_{mq}$ as the Washington DC-Mexico City distance. Using Google Map, we get $R_{mn} \approx 733$ km and $R_{mq} \approx 3032$ km. From these data, we get $\beta \approx 1.7$. For the sake of simplicity, in this work, we will take $\boldsymbol{\beta} \approx \mathbf{3/2}$ as an approximation, and the trade equation for the two countries becomes

$$Trade_{mn}(t) \approx \frac{1}{\omega_{mn}(t)} \frac{(E_m(t)I_n(t))^{0.47} + (I_m(t)E_n(t))^{0.47}}{R_{mn}^{3/2}} \quad \text{(s12)}$$

**The Relationship Between Export and GDP**

We assume that exports and GDP satisfy the relationship $E = kG^{\rho}$, or in the normalized variable, satisfy

$$\tilde{E} = k'\tilde{G}^{\rho} \quad \text{(s13)}$$

where $\tilde{E} = E/E_{max}$ and $\tilde{G} = G/G_{max}$. We will find the parameter $\rho$ for NAFTA and G7 member countries. Equation (21) can be changed in the form $\ln \tilde{E} = \ln k' + \rho \ln \tilde{G}$.

Figure s5 is the curve of $\ln \tilde{E}$ against $\ln \tilde{G}$ for NAFTA member countries. For USA and Mexico, we take data from 2000-2019; for Canada, we take data from 2009-2019. If we use data before 2009 for Canada, there is a high scatter behaviour. If the data is fitted with a linear curve, we get ($\rho = 1.33, R^2 = 0.886$), ($\rho = 1.28, R^2 = 0.899$), and ($\rho = 2.05, R^2 = 0.772$) for USA, Canada, and Mexico, respectively.

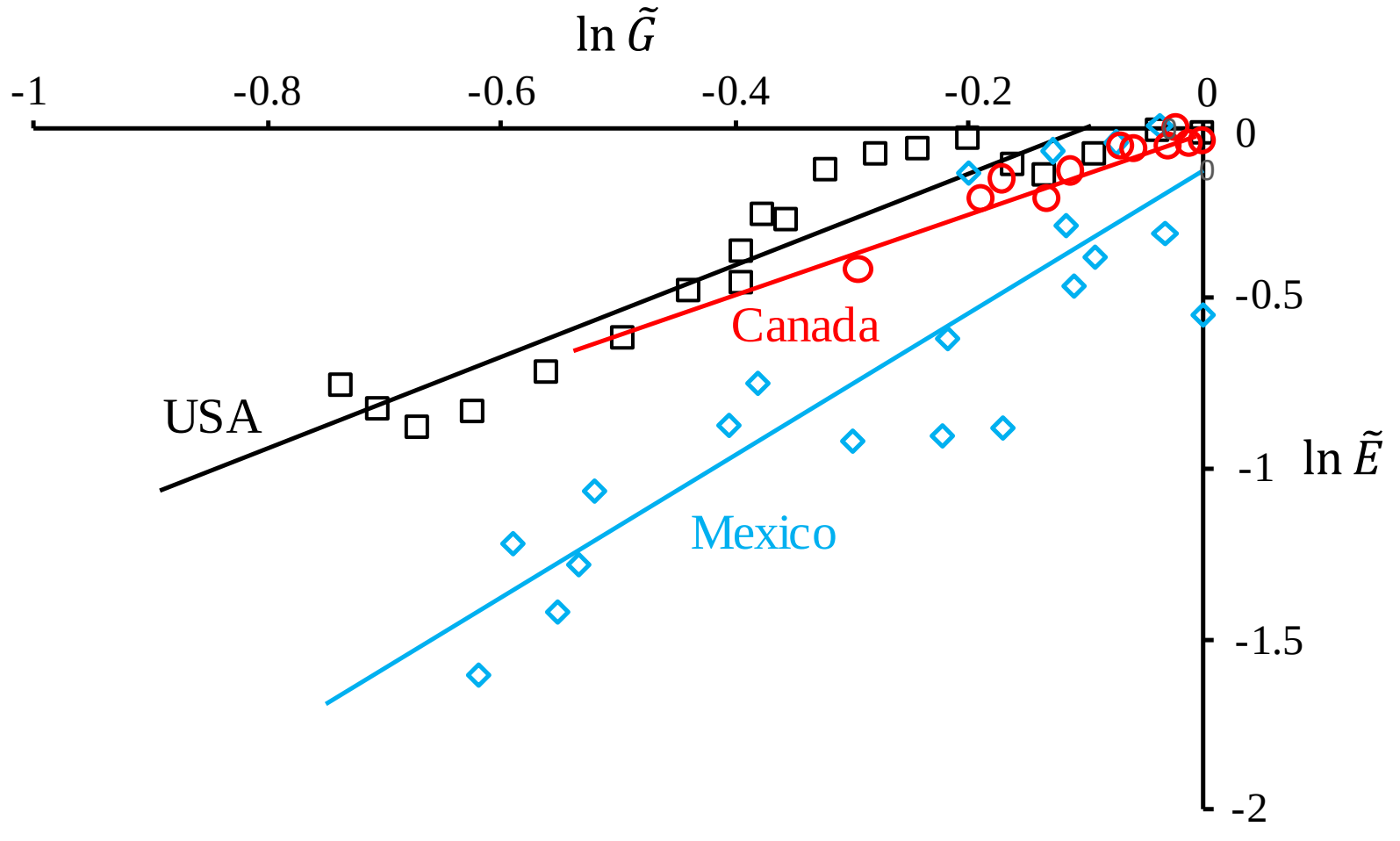

FIG. s5. Dependence of $\ln \tilde{E}$ against $\ln \tilde{G}$ for NAFTA member countries. Symbols (squares = USA, circles = Canada, diamonds = Mexico) are data from [3] while the lines are the result of linear fitting.

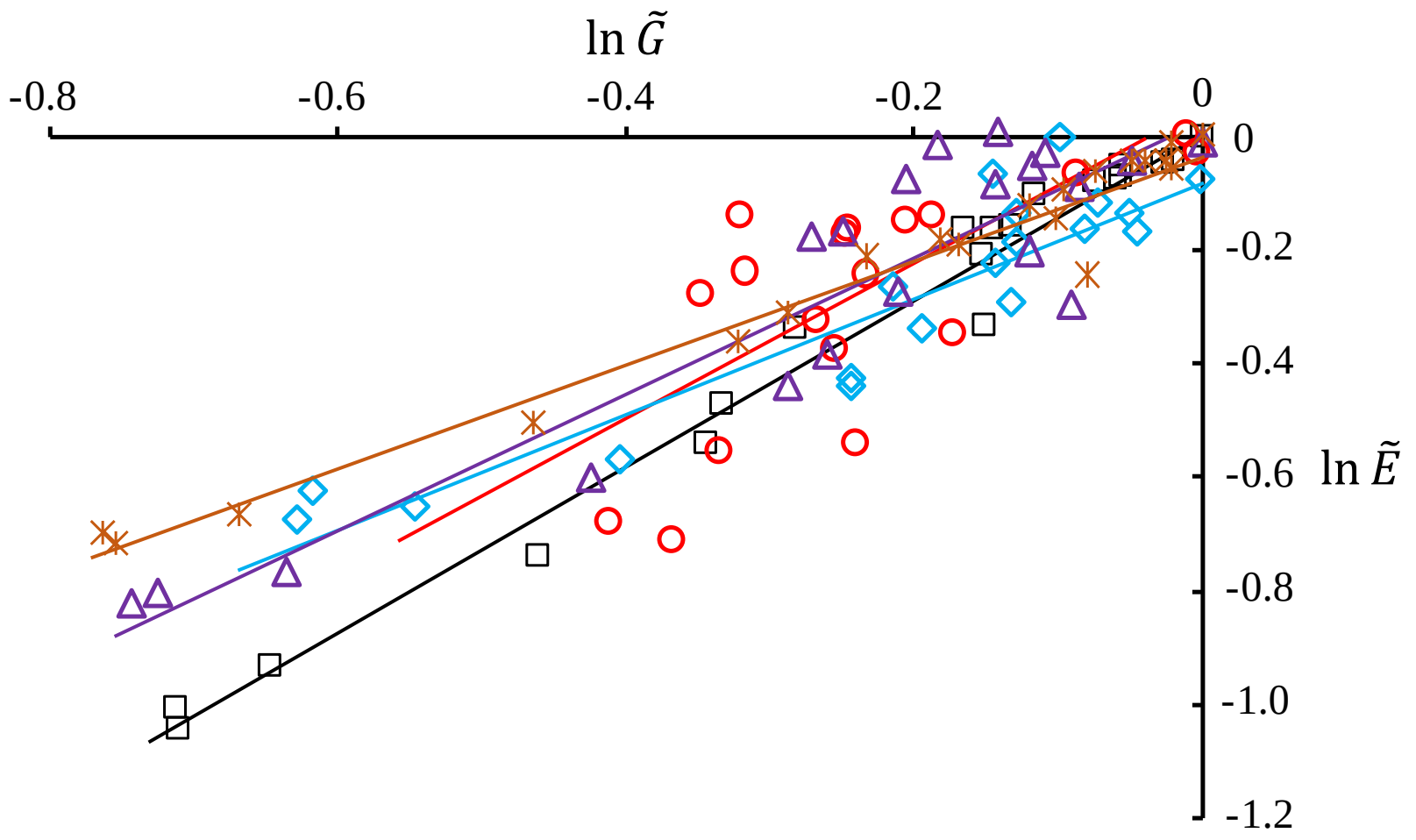

FIG. s6. Dependence of $\ln \tilde{E}$ against $\ln \tilde{G}$ for G7 member countries excluding USA and Canada. Symbol (squares = Germany, circles = Japan, diamonds = UK, triangles = Italy, stars = France) are data from [3] while the lines are the result of linear fitting.

The data for the G7 member countries except for the USA and Canada (because they have already been counted as NAFTA members) are shown in Fig. s6. We get the following parameters $(\rho = 1.48, R^2 = 0.9825)$, $(\rho = 1.39, R^2 = 0.504)$, $(\rho = 0.99, R^2 = 0.845)$, $(\rho = 1.20, R^2 = 0.860)$, and $(\rho = 0.93, R^2 = 0.968)$, for Germany, Japan, the UK, Italy, and France, respectively.

## The Relationship Between Export and Import

Next, we determine the relationship between exports and imports and whether it satisfies the linear equation. We use data from several G20 member countries: China, South Korea, Indonesia, and Brazil. Figure s7 is the curve of $\tilde{I} = I/I_{max}$ against $\tilde{E}$ for these countries. If the data group is fitted with a linear line, $\tilde{I} \propto \tilde{E}$, we get a fairly good fitting result where (the slope, $R^2$ value) are (0.97,0.985), (1.024,0.91), (0.97,0.922), dan (0.91,0.917) for China, South Korea, Indonesia, and Brazil, respectively. These results prove that imports and exports have a linear relationship.

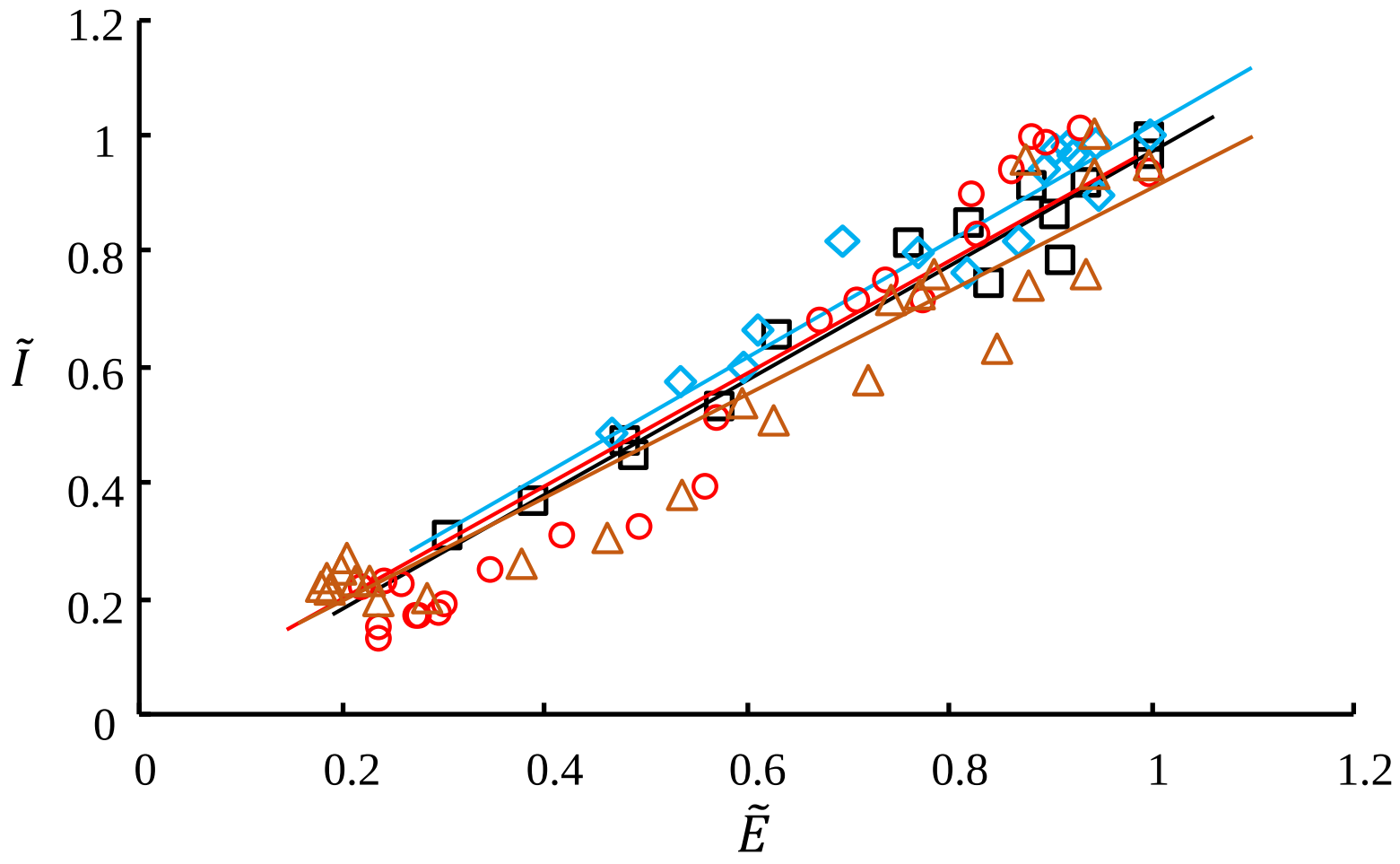

FIG. s7. The dependence of $\tilde{I} = I/I_{max}$ on $\tilde{E}$ for China (square), South Korea (diamonds), Indonesia (circles), and Brazil (triangle). Symbols are data from [3] while the lines are the result of linear fitting.

Based on Eq. (s13) and the results shown in Fig. s7, we can write $\tilde{I} = k"G^{\rho}$. Thus, Eq. (s7) can be written as

$$F_{mn}(t) = \frac{2k\prime k"}{\omega_{mn}(t)} \frac{G_m^{\delta} G_m^{\delta}}{R_{mn}^{\beta}} \qquad \text{(s14)}$$

with $\delta = \alpha\rho$. Suppose we use the average value of $\rho$ NAFTA and G7 member countries, we get $\langle\rho\rangle = 1.33$ and $\boldsymbol{\delta \approx 0.665}$. If we use $\boldsymbol{\beta \approx 3/2}$, Eq. (14) can be written as

$$F_{mn}(t) \approx \frac{2k\prime k"}{\omega_{mn}(t)} \frac{G_m^{\delta} G_n^{\delta}}{R_{mn}^{3/2}} = K \frac{G_m^{0.665} G_n^{0.665}}{T_{mn}^{3/2}} \qquad \text{(s15)}$$

where $K = 2k'k"$ and $T_{mn} = \omega_{mn} R_{mn}$.

**Table A1** Parameter $\alpha$ obtained from the pairs of countries analyzed (ASEAN, G7, NAFTA, Mercusor, and East Asia).

| Region | Country 1 | Country 2 | $\alpha$ |
|---|---|---|---|
| ASEAN | Indonesia | Malaysia | 0.52 |
| | | Singapore | 0.53 |
| | | Thailand | 0.65 |
| | | Philippines | 0.61 |
| | | Vietnam | 0.70 |
| | Malaysia | Singapore | 0.55 |
| | | Thailand | 0.44 |
| | | Philippines | 0.53 |
| | | Vietnam | 0.75 |
| | Singapore | Thailand | 0.33 |
| | | Philippines | 0.70 |
| | | Vietnam | 0.38 |
| | Thailand | Philippines | 0.56 |
| | | Vietnam | 0.61 |
| | Philippines | Vietnam | 0.60 |
| NAFTA | USA | Canada | 0.47 |
| | | Mexico | 0.55 |
| | Canada | Mexico | 0.73 |
| Mercosur | Brazil | Argentina | 0.52 |
| | | Uruguay | 0.42 |
| | | Paraguay | 0.34 |
| | Argentina | Uruguay | 0.32 |
| | | Paraguay | 0.37 |
| G7 | USA | Germany | 0.40 |
| | | UK | 0.32 |
| | | France | 0.40 |
| | | Italy | 0.42 |
| | | Canada | 0.35 |
| | Germany | UK | 0.39 |
| | | France | 0.40 |
| | | Italy | 0.37 |
| | UK | France | 0.24 |
| | France | Italy | 0.35 |
| East Asia | China | Japan | 0.39 |
| | | South Korea | 0.53 |
| | | Hongkong | 0.51 |
| | Japan | South Korea | 0.61 |
| Selected EU | Switzerland | Germany | 0.41 |
| | | France | 0.43 |
| | | Italy | 0.35 |

# Supplementary 2

The lognormal cumulative distribution function (CDF) can be expressed as a linear equation

$$\mathrm{erf}^{-1}(2CDF-1)=\frac{\ln G}{s\sqrt{2}}-\frac{\mu}{s\sqrt{2}} \qquad \text{(s16)}$$

where $\mathrm{erf}(x)$ denotes the error function and $\mathrm{erf}^{-1}(x)$ the inverse of the error function, $\mu$ is the mean, and $s$ represents the standard deviation. Figure s8 shows curves of the $\mathrm{erf}^{-1}(2CDF\text{ -}1)$ against $\ln G$ using data from World Bank data [World Bank, https://data.worldbank.org/indicator/NY.GDP.MKTP.CD, accessed 10 August 2022] to relevant results from Eq. (s16) for the years: (a) 2005, (b) 2010, (c) 2015, and (d) 2020. Based on the $R^2$ values we conclude that the resulting fittings are quite accurate. This finding supports the hypothesis that the GDP follows the lognormal distribution.

We calculate the parameter $s$ using the slope of the curve, $1/s\sqrt{2}$. For 2005, 2010, 2015, and 2020, the fitting results yield $s=2.653$, $s=2.637$, $s=2.613$, and $s=2.599$, respectively. Based on the intercepts, the means for 2005, 2010, 2015, and 2020 are $\mu=31.436$, $\mu=31.778$, $\mu=31.835$, and $\mu=31.837$, respectively. The mean appears to rise with increasing time. We get a slope of 0.0252 from the fitting results (Fig. s8(e)).

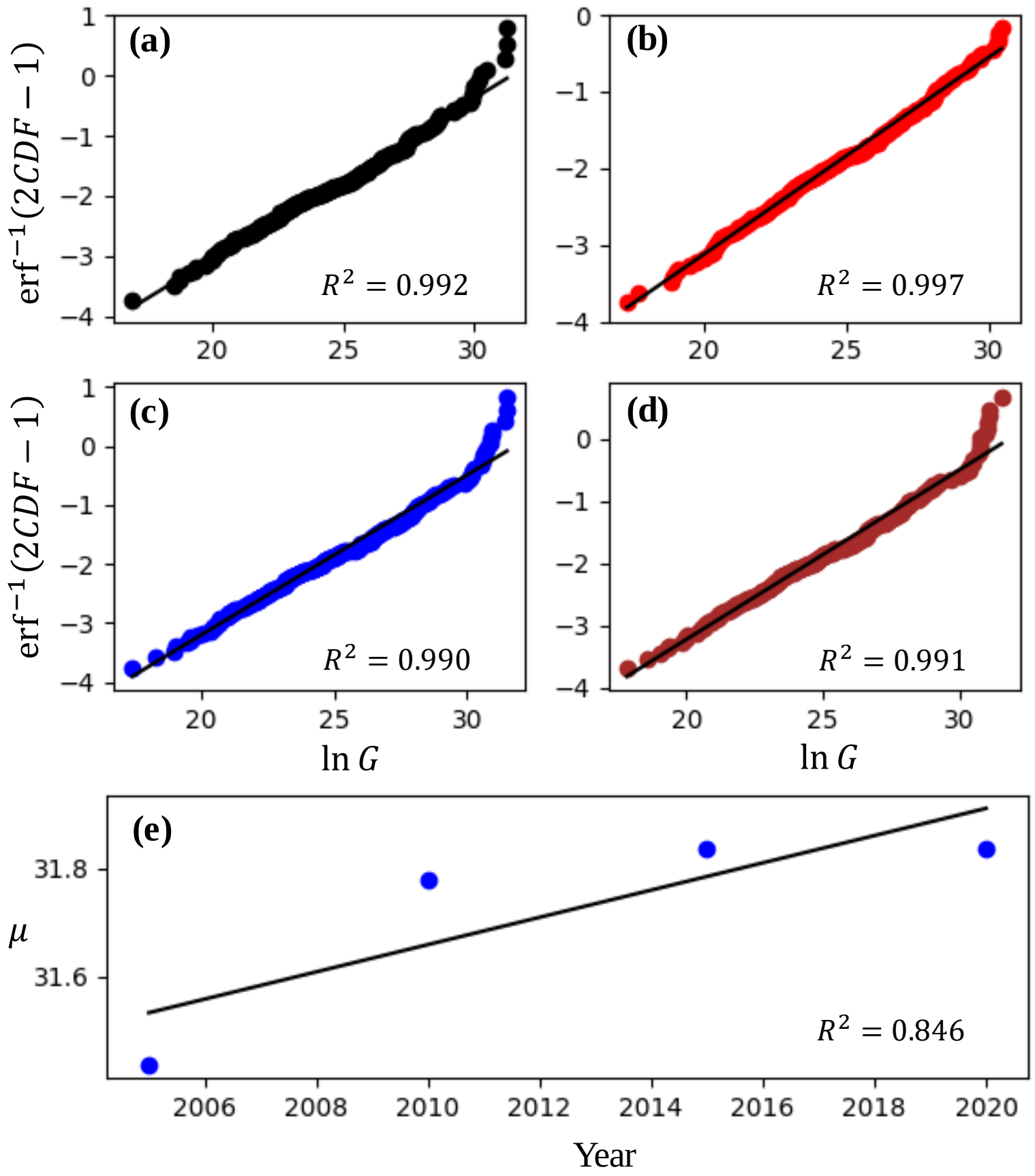

FIG. s8. (a)-(d) Fitting of erf$^{-1}$ ($2CDF$ -1) as a function of $\ln G$ using World Bank data [World Bank, https://data.worldbank.org/indicator/NY.GDP.MKTP.CD, accessed 10 August 2022] (symbols) and fitting results (lines) for the years: (a) 2005, (b) 2010, (c) 2015, and (d) 2020. (e) Variation in $\mu$ as a function of time: symbols represent data from fitting results in (a)-(d), and the line represents the linear fitting curve.

## Supplementary 3

$E[X^n] = \exp(n\mu + n^2 s^2/2)$ is one of the lognormal distribution function's properties so we can write

$$\int_0^\infty G^{4j\delta/3+2\delta/3} f(G)dG = \exp[(4j\delta/3 + 2\delta/3)\mu + (4j\delta/3 + 2\delta/3)^2 s^2/2] \qquad \text{(s17)}$$

We obtained $\delta \approx 0.665$ (see **Supplementary 1**), so equation (s17) becomes

$$\int_0^\infty G^{4j\delta/3+2\delta/3} f(G)dG \approx \exp[(0.89j + 0.44)\mu + (0.89j + 0.44)^2 s^2/2] \qquad \text{(s18)}$$

From the fitting in Fig.s8, we get $s \approx 2.6$ and $\mu \approx 31$ so that we can prove that up to $j = 9$, we get $(0.89j + 0.44)\mu > (0.89j + 0.44)^2 s^2/2$. The GDP growth rate, which is typically less than 0.1 (10%) [https://data.worldbank.org/indicator/NY.GDP.MKTP.KD.ZG, accessed 10 August 2022], can be used to approximate the $f$ value (see explanation below). Thus, $\left(\kappa f^2 4/g^{10/3}\right)^j$ decreases as $j$ increases. At the same time $1/j!$ decreases rapidly with increasing $j$. Thus, the terms in Eq. (18) are only significant at some of the initial $j$ values. At these values of $j$, we can roughly approximate $(0.89j + 0.44)\mu \gg (0.89j + 0.44)^2 s^2/2$ so that

$$\int_0^\infty G^{4j\delta/3+2\delta/3} f(G)dG \approx \exp[(4j\delta/3 + 2\delta/3)\mu] \qquad \text{(s19)}$$

Thus, equation (15) can be approximated as

$$w(g,f) \propto \frac{1}{g^{8/3}} \sum_{j=0}^{\infty} (-1)^j \frac{1}{j!} \left( \frac{\kappa}{4g^{\frac{10}{3}}} f^2 \right)^j \exp\left[ \left( \frac{4j\delta}{3} + \frac{2\delta}{3} \right) \mu \right]$$

$$= \frac{1}{g^{8/3}} e^{\frac{2\delta\mu}{3}} \sum_{j=0}^{\infty} (-1)^j \frac{1}{j!} \left( \frac{\kappa}{4g^{\frac{10}{3}}} f^2 e^{\left(\frac{4\delta\mu}{3}\right)} \right)^j$$

$$= \frac{1}{g^{8/3}} \exp\left( \frac{2\delta\mu}{3} - \frac{\kappa}{4g^{\frac{10}{3}}} f^2 e^{\left(\frac{4\delta\mu}{3}\right)} \right) \qquad \text{(s20)}$$

If we use $\delta = 0.665 \approx 2/3$ (see **Supplementary 1**) we obtain

$$w(g,f) \propto \frac{1}{g^{\frac{8}{3}}} \exp\left( \frac{4\mu}{9} - \frac{\kappa e^{8\mu/9}}{4g^{\frac{10}{3}}} f^2 \right) \qquad \text{(s21)}$$

Because the value of $\mu$ does not change significantly over the course of 20 years (**Fig. s8(e)**), we can write

$$w(g,f) = \frac{B}{g^{\frac{8}{3}}} \exp\left( -\frac{(f/\kappa')^2}{g^{\frac{10}{3}}} \right) \qquad \text{(s22)}$$

where $\kappa' = 2/\sqrt{\kappa e^{8\mu/9}}$ and $B$ is a constant.

## Supplementary 4:

The total $g$ experienced by a country is

$$g_j \propto \sum_{i\neq j} \frac{G_i^\delta}{\omega_{j,i} R_{j,i}^\beta} \tag{s23}$$

We have data on the condition that the distance between states is assumed to be constant (there is no change in the dielectric constant in the year that is our focus). From equation (s23) we can write

$$f_j = \frac{dg_j}{dt} \propto \delta \sum_{i\neq j} \frac{G_i^{\delta-1}\frac{dG_i}{dt}}{\omega_{j,i} R_{j,i}^\beta} \tag{s24}$$

$$\propto \delta \sum_{i\neq j} \frac{G_i^\delta}{\omega_{j,i} R_{j,i}^\beta}\left(\frac{1}{G_i}\frac{dG_i}{dt}\right) \tag{s25}$$

However, $(1/G_i)(dG_i/dt)$ is the GDP growth rate of country $i$ which will affect country $j$. We can assume that

$$p_{j,i} \propto \frac{G_i^\delta}{\omega_{j,i} R_{j,i}^\beta} \tag{s26}$$

is the weight that the i-th country to $f_j$ so we can write

$$f_j \propto \sum_{i\neq j} p_{j,i}\left(\frac{1}{G_i}\frac{dG_i}{dt}\right) \tag{s27}$$

We assume $p_{j,j} = 1$ so that equation (s26) can be written as

$$f_j \propto \left[\frac{1}{G_j}\frac{dG_j}{dt} + \sum_i \; p_{j,i}\left(\frac{1}{G_i}\frac{dG_i}{dt}\right)\right] \tag{s28}$$

If we assume that the total change in GDP multiplied by the weight is close to zero or very small compared to $(1/G_j)(dG_j/dt)$ (indeed this assumption still needs confirmation), we get

$$f_j \propto \frac{1}{G_j}\frac{dG_j}{dt} \tag{s29}$$

So, the $f$ value of a country is proportional to the country's GDP growth.